\documentstyle[aps,pra,epsfig,rotate,twocolumn]{revtex}
\def\A{{\bf A}_{\rm i}}
\def\gta{\mathrel{{\lower 3pt\hbox{$\mathchar"218$}}\hskip-7pt
   \raise 2pt\hbox{$>$}}}
\def\be{\begin{equation}}
\def\ee{\end{equation}}
\begin{document}
\title{Spontaneous breaking of axial symmetry for Schr\"odinger's
equation in the presence of a magnetic field}
\author{Jorge Berger}
\address{Department of Sciences, Ort Braude College, P.O. Box 78,
21982 Karmiel, Israel and \\
Department of Physics, Technion, 32000 Haifa, Israel}
\flushbottom
\twocolumn[
\hsize\textwidth\columnwidth\hsize\csname @twocolumnfalse\endcsname
\maketitle
\begin{abstract}
For appropriate parameters, the ground state for the Schr\"odinger
and Amp\`ere coupled equations in a cylindric domain does not have
axial symmetry.
\end{abstract}
\pacs{}   ]

Symmetry breaking --- a situation in which the solution of a problem
has lower symmetry than the problem itself --- is usually a room for
interesting physics (e.g. \cite{Kib}). It is therefore gratifying to note
that symmetry 
breaking is found in one of the most common problems we encounter:
that of a condensate of particles interacting with a magnetic field,
in a time independent state. 

Our model is as follows. We define a ``thermodynamic potential" density
\be
g=|\nabla\times\A|^2+|(i\nabla-\A-br\hat\theta)\psi|^2 \; ,
\label{g}
\ee
where $\psi$ is a wave function, $\A$ the magnetic potential induced
by $\psi$, $b$ is the external magnetic field, which is taken as
uniform, and $r,\theta$ are cylindrical coordinates, with the $z$-axis
parallel to the field. $b$ and $\A$ are normalized so that no
coefficients are required in (\ref{g}). The sample in which electrical
currents can exist will be an infinitely long cylinder, and therefore
$\nabla\times\A=0$ outside of the sample. The first term in (\ref{g})
gives the contribution of the magnetic field to the thermodynamic
potential, subject to the constraint of a fixed external field; the
second term may be associated to the kinetic energy and provides 
for the interaction between the magnetic field and
the particles described by $\psi$. Our model has 2 fixed parameters:
$b$ and the average density of particles, which is defined by
\be
\rho=\frac{1}{\pi}\int |\psi|^2 dS \; ,
\label{ro}
\ee
where $\int dS$ denotes the integral over the cross section of the 
cylindrical sample; this cross section will be taken as a disk and its
radius as the unit of length. (If the radius is denoted by $R$, $\psi$
and $\A$ scale as $R^{-1}$, $b$ and $\mu$ as $R^{-2}$, and $\rho$
as $R^{0}$.)

Our problem is to find the fields $\psi$ and $\A$ which minimize
the thermodynamic potential
$\int gdS$ for given $b$ and $\rho$. For this purpose we minimize
$\int (g-\mu|\psi|^2)dS$, where $\mu$ is a Lagrange multiplier.
Variation of $\psi$ gives
\be
(i\nabla-\A-br\hat\theta)^2\psi=\mu\psi \; ,
\label{Sch}
\ee
which is the usual time independent equation for a charged particle in 
a magnetic field. Here, instead of a single particle we consider a
``condensate"; by this we mean that instead of a single particle there
are $\rho$ particles per unit volume, but they don't interact among
themselves and they are all in the same state. The condensate obeys the
same equation as a single particle, but its current and magnetic influence
will be proportional to $\rho$. In the limit $\rho\rightarrow 0$, 
the thermodynamic potential approaches $\pi\mu\rho$ and Eq.~(\ref{Sch}) 
reduces to the Landau problem \cite{Landau}.

Variation of $\A$ gives
\be
\nabla\times\nabla\times\A={\mathrm Re}[\bar\psi(i\nabla-\A-br\hat\theta)\psi] \; ,
\label{Amp}
\ee
where the bar denotes complex conjugation. This is just Amp\`ere's law.
If $\A$ (resp. $\psi$) is fixed, then Eq.~(\ref{Sch}) (resp. (\ref{Amp}))
is linear, but the system of both equations is nonlinear due to their
mutual interaction.
We choose a gauge such that $\A$ is parallel to $\hat\theta$. The boundary
condition for $\A$ is continuity of $\nabla\times\A$. For $\psi$ we will
take the natural condition that its normal derivative vanishes at the
boundary.

There are solutions of the system (\ref{Sch})-(\ref{Amp}) with the 
axial symmetry of the problem. These have the form
\begin{eqnarray}
\psi&=&{\mathcal R}_m(r)e^{-mi\theta} \; , \nonumber \\
\A&=&{\mathcal A}_m(r)\hat\theta \; ,
\label{cyl}
\end{eqnarray}
where $m$ is an integer. For this form, (\ref{Sch})-(\ref{Amp}) reduces
to a system of ordinary differential equations. The value of $m$ has to
be chosen such that the minimum value of the thermodynamic potential is
obtained. This value of $m$ is an increasing function of $b$. Due to
continuity, $\psi$ has to vanish along the axis of the sample for $m\neq 0$.

We now ask whether there exist situations such that the minimizer of 
$\int gdS$ is not in the family (\ref{cyl}). In the following, we will no
longer consider $\rho$ and $b$ as independent parameters, but will focus
on the value of the magnetic field for which the lowest value of $\int gdS$
among the solutions in the family (\ref{cyl}) is shared by the winding
numbers $m=0$ and $m=1$. For $\rho\le 10$, it is found numerically that
this approximately occurs at $b=1.924+0.171\rho+0.00104\rho^2-0.000036\rho^3$.
We solve the problem in two stages: in
the first stage we perform a variation in which $\psi$ has the form
$p+qr(1-r/2)e^{-i\theta}$ and 
$A_{{\rm i}\theta}=a_0(r-2r^2/3)+a_1(r^2-3r^3/4)\cos\theta$.
This variation can be performed analytically, enabling 
us to obtain a qualitative picture of the minima and saddle points of
$\int gdS$. After we know what to look for, the system (\ref{Sch})-(\ref{Amp})
is solved numerically.

\begin{figure}[t]
\begin{center}
\includegraphics[width=0.4\textwidth]{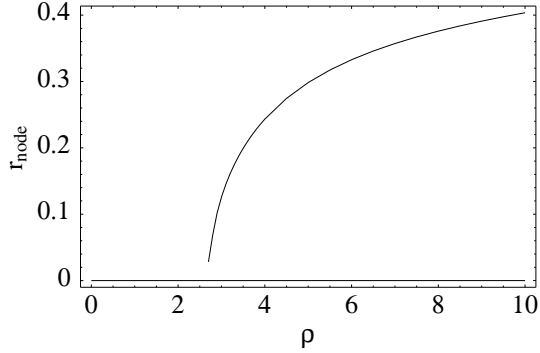}
\end{center}
\caption[]{Distance of the nodal line, $r_{\rm node}$, from the axis of the
sample (divided by the sample radius). For a solution in the family (\ref{cyl})
with $m=1$, the nodal line is at the axis of the sample. At $\rho\sim 2.7$, a
new (asymmetric) solution bifurcates from it.}
\label{vort}
\end{figure}

Besides the symmetric solutions with $m=0$ or $m=1$, we always find a
nonsymmetric solution, but this has a higher value of $\int gdS$. 
(By ``a solution", we mean a class of equivalent solutions.) However,
for $\rho\gta 2.7$, a bifurcation occurs from the symmetric solution with
$m=1$. This bifurcation is characterized by a migration of the nodal line 
away from the axis of the sample (Fig.~\ref{vort}). This behavior reminds
of a transition numerically found for mesoscopic superconducting disks
\cite{Peeters}, except that in the present situation there is a single
nodal line.

\begin{figure}[bt]
\begin{center}
\begin{tabular}{rl}
\protect{\put(5,0){\begin{rotate}
{$~~~~~~~~~~~~\frac{1}{\rho}\int(g-g_{\rm sym})dS$}\end{rotate}} }
&
\includegraphics[width=.4\textwidth]{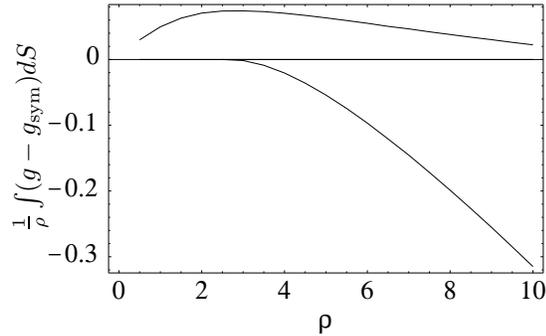}
\end{tabular}
\end{center}
\caption[]{Thermodynamic potential of asymmetric solutions, in comparison to 
those of symmetric solutions. $g_{\rm sym}$ is the potential density that gives 
the lowest integral $\int g_{\rm sym}dS$ in the symmetric family with $m=0$ or 
$m=1$. (For the magnetic field considered, $m=0$ and $m=1$
give the same integral.) For $\rho\gta 2.7$, the lowest potential is obtained
for an asymmetric solution.}
\label{energy}
\end{figure}

We find that this nonsymmetric solution has lower thermodynamical potential 
than any couple of fields in the family (\ref{cyl}) (Fig.~\ref{energy}).
This means that the minimizer of $\int gdS$ is not in the symmetric
family (\ref{cyl}), thus giving rise to symmetry breaking.

One might suspect that the solutions of (\ref{Sch})-(\ref{Amp}) we have found 
are not those with the lowest thermodynamical potential. This is unlikely.
For $\rho\rightarrow 0$, our symmetric solutions approach well known analytic 
results; for $\rho=10$ we have solved (\ref{Sch})-(\ref{Amp}) for initial 
symmetric configurations in a broad range, and none of the trials lead to
a lower potential. We might have missed a non-symmetric solution with
lower potential, but this possibility would only strengthen our conclusion
that axial symmetry is broken.


\begin{references}
\bibitem{Kib}D. A. Kirzhnits, ZhETF Pis. Red. {\bf 15}, 745 (1972) [JETP
Lett. {\bf 15}, 529 (1972)].
\bibitem{Landau}R. B. Dingle, Proc. R. Soc. London, Ser. A {\bf 211}, 500 
(1952).
\bibitem{Peeters}V. A. Schweigert, F. M. Peeters, and P. S. Deo, Phys. Rev.
Lett. {\bf 81}, 2783 (1998).
\end{references}
\end{document}